\documentclass[aps,pra,twocolumn,showpacs,superscriptaddress]{revtex4-1}
\usepackage{amsmath,amssymb}
\usepackage[matha,mathx]{mathabx}
\usepackage{txfonts}
\usepackage{float}
\usepackage[pdftex]{graphicx}
\usepackage{subfigure}
\usepackage{natbib}
\usepackage{bm}
\newcommand \beq{\begin{eqnarray}}
\newcommand \eeq{\end{eqnarray}}

\begin{document}

\title{Two-leg fermionic Hubbard ladder system in the presence of
state-dependent hopping}
\author{Shun Uchino}
\affiliation{DQMP, University of Geneva, 24 Quai Ernest-Ansermet,
1211 Geneva, Switzerland}
\author{Thierry Giamarchi}
\affiliation{DQMP, University of Geneva, 24 Quai Ernest-Ansermet,
1211 Geneva, Switzerland}

\date{\today}

\begin{abstract}
We study a two-leg fermionic Hubbard ladder model with
a state-dependent hopping. We find that, contrary to the case without 
a state-dependent hopping, for which
the system has a superfluid nature regardless of the sign of the
interaction at incommensurate filling, 
in the presence of such a hopping a spin-triplet superfluid, 
spin-density wave 
and charge-density wave phases emerge. We examine our results in the light of 
recent experiments on periodically-driven optical lattices in cold atoms.
We give protocols allowing us to realize the spin-triplet superfluid
elusive in the cold atoms. 
\end{abstract}


\maketitle

\section{Introduction}
Strongly correlated one-dimensional systems have attracted strong attention
over the past decades.
In general, in such systems 
the excitations differ strongly from their higher dimensional
counterparts and for fermions 
are very different from the usual Landau quasiparticles 
occurring in a Fermi-liquid state \cite{nozieres1964theory}.
Instead, many of the one dimensional systems belong to
the universality class known to be the Tomonaga-Luttinger liquid 
\cite{giamarchi2003quantum}.

In particular, the system made of two coupled fermionic chains, namely,
the two-leg ladder system, has been intensively studied in the past
\cite{PhysRevB.47.10461,PhysRevB.48.15838,PhysRevB.50.252,
nagaosa1995bipolaron,PhysRevB.53.R2959,PhysRevB.53.12133,PhysRevB.53.14036,
PhysRevB.53.11721,noack1996ground,PhysRevB.65.165122,PhysRevB.75.245119,giamarchi2003quantum}.
This system has been shown to exhibit 
superconductivity, not only for attractive 
interactions ($s$-wave superconductivity), but also quite remarkable 
for purely repulsive ones. 
In the latter case the superconductivity is of $d$-wave symmetry. 
The $d$-wave superconductivity emerges by doping of a Mott insulating phase at
half filling. 

While the one-dimensional system has been intensively studied 
as a first step towards other materials in
higher dimensions, such as the high-$T_c$ superconductors, 
nowadays it is a major subject in itself due to the 
relevance for some experiments, 
in particular in the field of cold atomic gases \cite{RevModPhys.80.885}.

Indeed, due to rapid advances in technology, 
cold atoms are a promising way to investigate the one
 dimensional systems with an unprecedented level 
of control on the interchain hopping and interactions. 
Most of the atoms utilized in experiments have
internal degrees of freedom, which correspond to hyperfine states
when we focus on alkali species, already 
allowing to reproduction of models 
such as the 
Hubbard model \cite{greiner2002collapse,PhysRevLett.94.080403}. 
More recently, ladder systems have also been produced, 
both for bosonic and fermionic states 
\cite{PhysRevA.73.033605,folling2007direct,chen2011many,greif2013short,atala2014observation}.

In addition to simulating systems directly existing in condensed matter
physics, by using the unique manipulations available in experiments,
 cold atoms also allow us to realize new quantum states of matter.

One such extension, which is the focus of this paper,
is the time modulation of optical lattices
\cite{PhysRevLett.67.516,PhysRevLett.95.260404,
PhysRevLett.99.220403,struck2011quantum,PhysRevX.4.031027}.
By applying such a modulation with sufficiently high frequencies,
it is possible to tune the hopping matrix.
This technique allows one to control
the hopping not just in strength but also
in sign, since the renormalized hopping is essentially
proportional to a Bessel function.
In addition, by using the state-dependent optical lattice
\cite{PhysRevLett.91.010407,mandel2003controlled}
or applying a magnetic field one can also control the hopping matrix
element in a state-dependent manner.
In fact,  
such a setup has motivated several theoretical studies on
existence or non-existence of exotic paired states in the two-dimensional
Hubbard model \cite{PhysRevA.70.033603,PhysRevLett.103.025303,
PhysRevB.83.115104,2014arXiv1408.3119G}, 
and on the
presence of incommensurate density waves and segregation
in the one-dimensional Hubbard model
\cite{PhysRevLett.95.226402,PhysRevB.81.224512}.

One may also expect the realization of an unconventional superfluid   
in cold atoms by means of such a unique technique.
To realize a superfluid in cold atoms, so far,
it is necessary to use a 
Feshbach resonance, since the typical temperature in the experiments
is of the order of a tenth of the Fermi temperature  \cite{RevModPhys.80.885}. 
A weak-coupling BCS transition 
temperature is extremely low compared to this temperature. 
A Feshbach resonance allows one to boost 
the interactions enough so that $s$-wave superfluidity 
can be routinely realized for attractive
interactions. However, other symmetries are not so easily attainable. 
A $p$-wave Feshbach resonance is unstable due to the atom-molecule
and molecule-molecule inelastic collisions \cite{RevModPhys.82.1225}.
Therefore, the realization of an unconventional superfluid with cold atoms is
a highly challenging issue.

In this paper, we show how one can realize a spin-triplet superfluid in a 
two-leg Hubbard ladder system. 
In the presence of a state-dependent hopping, the $d$-wave pairing state
in the normal ladder is replaced 
by a spin triplet superfluid and a spin density wave (SDW) state. 
We also discuss the case of an attractive interaction which would lead
in the absence of state dependent hopping to $s$-wave superconductivity and 
which gives an incommensurate charge density wave (CDW) in the presence of 
state-dependent hopping. 

With a ladder system we thus show that we can obtain a spin-triplet state with 
purely local ($s$-wave) 
repulsive interactions, which is an attainable situation in
experiments. 
In a single chain such a state would have demanded an extended Hubbard model 
with on-site repulsion and nearest-neighbor attraction
of the same order of magnitude \cite{PhysRevB.49.9670}, something which is at 
the moment out of reach in cold atomic systems. 

This paper is organized as follows. Section \ref{sec:hamiltonian} 
discusses the Hamiltonian we propose and its low-energy description
by means of the bosonization technique.
In Sec. \ref{sec:rg}, the possible phases are determined
by using a renormalization group analysis.
In Sec. \ref{sec:discussion}, we discuss
the properties of the strong-coupling limit in the system
and experimental protocols toward its realization. 
Section \ref{sec:conclusion} is the Conclusion. 
Technical details can be found in the Appendix. 

\section{Hamiltonian}
\label{sec:hamiltonian}

\begin{figure}[h]
 \begin{center}
  \includegraphics[width=1\linewidth]{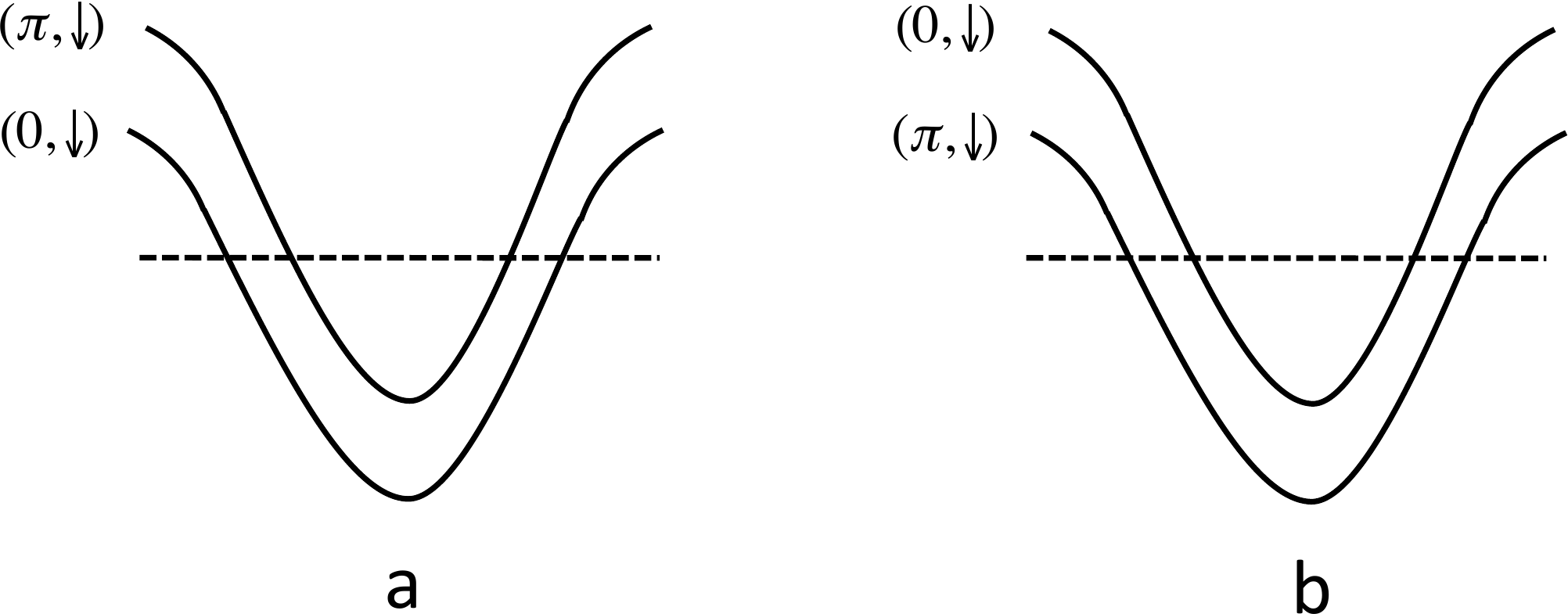}
  \caption{Band structure of the two-leg fermionic Hubbard ladder of
atoms with spin down
(a) without the state-dependent hopping and
(b) with state-dependent hopping as
  $t_{\perp\uparrow}=-t_{\perp\downarrow}$.
In each case, there are four different points at the Fermi level.
If the repulsive interaction is added, the latter leads to
a spin-triplet superfluid while the former leads to a $d$-wave superfluid
at incommensurate filling.
The band structure of atoms with spin up does not change
in the presence of the state-dependent hopping.
}
  \label{fig1}
 \end{center}
\end{figure}
We study two-component fermions confined in
the two-leg ladder geometry.
Our starting point is the following two-leg Hubbard ladder model:
\beq
H
&=&-t_{\parallel}\sum_{j=1}^N\sum_{s=\uparrow,\downarrow}
\sum_{p=\pm1}(
c^{\dagger}_{j,s, p}c_{j+1,s, p}+h.c.)\nonumber\\
&&-\sum_{j,s}t_{\perp s}(c^{\dagger}_{j,s, 1}c_{j,s, -1}+h.c.)
+U\sum_{j,p}n_{j,\uparrow, p}n_{j,\downarrow, p},
\label{eq:original-h}
\eeq
where $t_{\parallel}$ and $t_{\perp s}$ are respectively
the hopping matrices
along the chain and rung directions,
and $j$ and $p$ indicate the chain and ladder indices.
Here, the on-site Hubbard $U$ can correspond to both
repulsive and attractive interactions, which indeed can be
realized experimentally.
We focus on a system at incommensurate filling
since we are interested in the stability of the superfluids
in the presence of the state-dependent hopping,
in particular, in the presence of such a hopping along the rung direction.
The effect of the state-dependent chain hopping has been partially
discussed in Refs.
\cite{PhysRevLett.95.226402,PhysRevLett.103.025303,PhysRevB.83.115104}.
In this section and section \ref{sec:rg}, we discuss the weak-coupling limit to
analyze the possible phases using a field theory analysis.
In our model, this condition implies
$t_{\parallel}\gg |U|,\ t_{\perp s}$.

To deal with the system in the weak-coupling limit correctly,
we first move to the bonding and anti-bonding representation for
the fermion operators:
\beq
c_{j,s, 0(\pi)}=[c_{j,s,1}+(-)c_{j,s,-1}]/\sqrt{2},
\eeq
which allows to diagonalize the hopping terms.
While in the absence of the rung hopping, the bonding and anti-bonding bands
are energetically degenerate,
these are split in the presence of the the rung hopping.
In the absence of the state-dependent rung hopping,
the splitting is independent of the states (or spins),
and therefore, there are four different points at the Fermi level
as can be seen from Fig.~\ref{fig1}.
In the presence of the state-dependent rung hopping,
however, the splitting starts to depend on the states
and leads to eight different points at the Fermi level.
At the same time, at $t_{\perp\uparrow}=-t_{\perp\downarrow}$,
the four point structure at the Fermi level is recovered
even though in this case the degeneracies occur between
$(\pi,\uparrow)$ and $(0,\downarrow)$ and between
$(0,\uparrow)$ and $(\pi,\downarrow)$ (see Fig. \ref{fig1}).
Then, the interaction term plays the role of
hybridization between the bonding and anti-bonding bands,
which is essential to lead to nontrivial states of matter
in the system.

We now consider the continuum limit to use
the bosonization.
The fermion in the continuum limit $\psi$ can be expressed
with conjugate phase fields $\phi$ and $\theta$ as
\cite{giamarchi2003quantum}
\beq
\psi_{s q r}(x)=\frac{1}{\sqrt{2\pi\alpha}}\eta_{s q}e^{irk_Fx}
e^{-i[r\phi_{s q}(x)-\theta_{s q}(x)]},
\label{eq:fermi-boson}
\eeq
with the Fermi momentum $k_F$, index $q=0$ or $\pi$ for
the bonding and anti-bonding bands,
index $r=-1 \ \text{or}\ 1$
for the left or right mover,
cut-off parameter $\alpha$, and
the phase fields $\phi_{s q}$ and $\theta_{s q}$
to be conjugate.
Here, we explicitly introduce
the Klein factor $\eta$, which
guarantees the correct anti-commutation relation of the fermions
and is also important to obtain correct expressions for
the bosonized Hamiltonian and correlation functions.
By substituting \eqref{eq:fermi-boson} into \eqref{eq:original-h},
one may obtain the following low-energy effective Hamiltonian:
\begin{widetext}
\beq
H=
\sum_{\mu=\rho,\sigma}\sum_{\nu=\pm}\int \frac{dx}{2\pi}
\left[u_{\mu \nu}K_{\mu \nu}(\nabla\theta_{\mu \nu})^2+
\frac{u_{\mu \nu}}{K_{\mu \nu}}(\nabla\phi_{\mu \nu})^2\right]
+\int \frac{dx}{2(\pi\alpha)^2}
[\cos2\phi_{\sigma+}\{g_1\cos(2\phi_{\sigma-}-\delta_{\sigma-}x)
+g_2\cos(2\phi_{\rho-}-\delta_{\rho-}x)\}\nonumber\\
+\cos2\theta_{\rho-}\{g_3\cos(2\phi_{\sigma-}-\delta_{\sigma-}x)
+g_4\cos2\phi_{\sigma+}\}
-\cos2\theta_{\sigma-}\{g_5\cos(2\phi_{\rho-}-\delta_{\rho-}x)+
g_6\cos2\phi_{\sigma+}\}],
\nonumber\\
\label{eq:boson}
\eeq
\end{widetext}
where
we introduced for $\phi$ fields,
\beq
\phi_{\rho+}=\frac{1}{2}
(\phi_{\uparrow0}+\phi_{\downarrow0}+
\phi_{\uparrow\pi}+\phi_{\downarrow\pi}),\\
\phi_{\rho-}=\frac{1}{2}
(\phi_{\uparrow0}+\phi_{\downarrow0}-
\phi_{\uparrow\pi}-\phi_{\downarrow\pi}),\\
\phi_{\sigma+}=\frac{1}{2}
(\phi_{\uparrow0}-\phi_{\downarrow0}+
\phi_{\uparrow\pi}-\phi_{\downarrow\pi}),\\
\phi_{\sigma-}=\frac{1}{2}
(\phi_{\uparrow0}-\phi_{\downarrow0}-
\phi_{\uparrow\pi}+\phi_{\downarrow\pi}),
\eeq
and similar relations for $\theta$ fields.
To obtain the above, we neglect the umklapp scatterings
since the system at incommensurate filling is concerned.
For our original Hamiltonian, we find
$\delta_{\rho-}=2K_{\rho-}(t_{\perp\uparrow}+t_{\perp\downarrow})/u_{\rho-}$,
$\delta_{\sigma-}=2K_{\sigma-}(
t_{\perp\uparrow}-t_{\perp\downarrow})/u_{\sigma-}$,
$g_i=U$ $(i=1, 2, \cdots,6)$.
In addition, $u_{\mu \nu}$ and $K_{\mu \nu}$ are the velocity and
the Tomonaga-Luttinger parameter, respectively.
We also note that to obtain the above bosonized Hamiltonian
\eqref{eq:boson},
we adopt
the following convention on the ordering of the Klein factors:
\beq
\eta_{\uparrow0}\eta_{\downarrow0}\eta_{\downarrow\pi}\eta_{\uparrow\pi}=1.
\eeq

\section{Renormalization group analysis}
\label{sec:rg}
Based on the bosonized Hamiltonian \eqref{eq:boson},
we now determine the possible phases in this model.
To this end, we employ the renormalization group (RG) approach
in the bosonized Hamiltonian \cite{giamarchi2003quantum}.
By performing the scaling of the cut-off $(\alpha\to\alpha'=\alpha
e^{dl})$,
one may obtain the set of the RG equations at the one-loop level (quadratic
with respect to the coupling constants), which is given by
(see Appendix)
\beq
&&\frac{dK_{\sigma-}}{dl}=-\frac{K_{\sigma-}^2J_0(\delta_{\sigma-}\alpha)
[y_1^2+y_3^2]}{2}+\frac{J_0(\delta_{\rho-}\alpha)y_5^2+y_6^2}{2},\\
&&\frac{dK_{\sigma+}}{dl}=-\frac{K_{\sigma+}^2[
J_0(\delta_{\sigma-}\alpha)y_1^2+J_0(\delta_{\rho-}\alpha)y_2^2+
y_4^2+y_6^2]}{2},\\
&&\frac{dK_{\rho-}}{dl}=-\frac{K_{\rho-}^2J_0(\delta_{\rho-}\alpha)
[y_2^2+y_5^2]}{2}+\frac{J_0(\delta_{\sigma-}\alpha)y_3^2+y_4^2}{2},\\
&&\frac{dy_1}{dl}=(2-K_{\sigma-}-K_{\sigma+})y_1-y_3y_4,\\
&&\frac{dy_2}{dl}=(2-K_{\rho-}-K_{\sigma+})y_2-y_5y_6,\\
&&\frac{dy_3}{dl}=(2-K_{\sigma-}-1/K_{\rho-})y_3-y_1y_4,\\
&&\frac{dy_4}{dl}=(2-K_{\sigma+}-1/K_{\rho-})y_4-y_1y_3J_0(\delta_{\sigma-}
\alpha),\\
&&\frac{dy_5}{dl}=(2-K_{\rho-}-1/K_{\sigma-})y_5-y_2y_6,\\
&&\frac{dy_6}{dl}=(2-K_{\sigma+}-1/K_{\sigma-})y_6-y_2y_5J_0(\delta_{\rho-}
\alpha),\\
&&\frac{d\delta_{\sigma-}}{dl}=\delta_{\sigma-}-\frac{K_{\sigma-}
J_1(\delta_{\sigma-}\alpha)[y_1^2+y_3^2]}{\alpha},\\
&&\frac{d\delta_{\rho-}}{dl}=\delta_{\rho-}-\frac{K_{\rho-}
J_1(\delta_{\rho-}\alpha)[y_2^2+y_5^2]}{\alpha},
\eeq
where the initial values are given as
$y_{i}(0)=U/(2\pi v_F)$ $(i=1,2,\cdots,6)$,
$K_{\rho-}(0)=K_{\sigma-}(0)=1$, $K_{\rho+}=1/\sqrt{1+U/(2\pi v_F)}$,
$K_{\sigma+}(0)=1/\sqrt{1-U/(2\pi v_F)}$
with the Fermi velocity $v_F$.
We note that since there is no cosine term with respect to
$\phi_{\rho+}$ and $\theta_{\rho+}$, which are
decoupled from the other  phase fields,
$K_{\rho+}$ does not flow up to this order of approximation.
In addition,
$J_{n}$ $(n=0,1)$ is the $n$th order Bessel function,
which plays a role in controlling the relevance 
of the corresponding cosine terms.
Thus, one may classify the fixed points into the following three
cases:
\beq
&&\text{(a)}\  \delta_{\rho-}\to\infty, \ \delta_{\sigma-}\to0,\nonumber\\
&&\text{(b)}\ \delta_{\rho-}\to\infty, \ \delta_{\sigma-}\to\infty,\nonumber\\
&&\text{(c)}\ \delta_{\rho-}\to0,\ \delta_{\sigma-}\to\infty.\nonumber
\eeq

\begin{figure}[h]
 \begin{center}
  \includegraphics[width=1.2\linewidth]{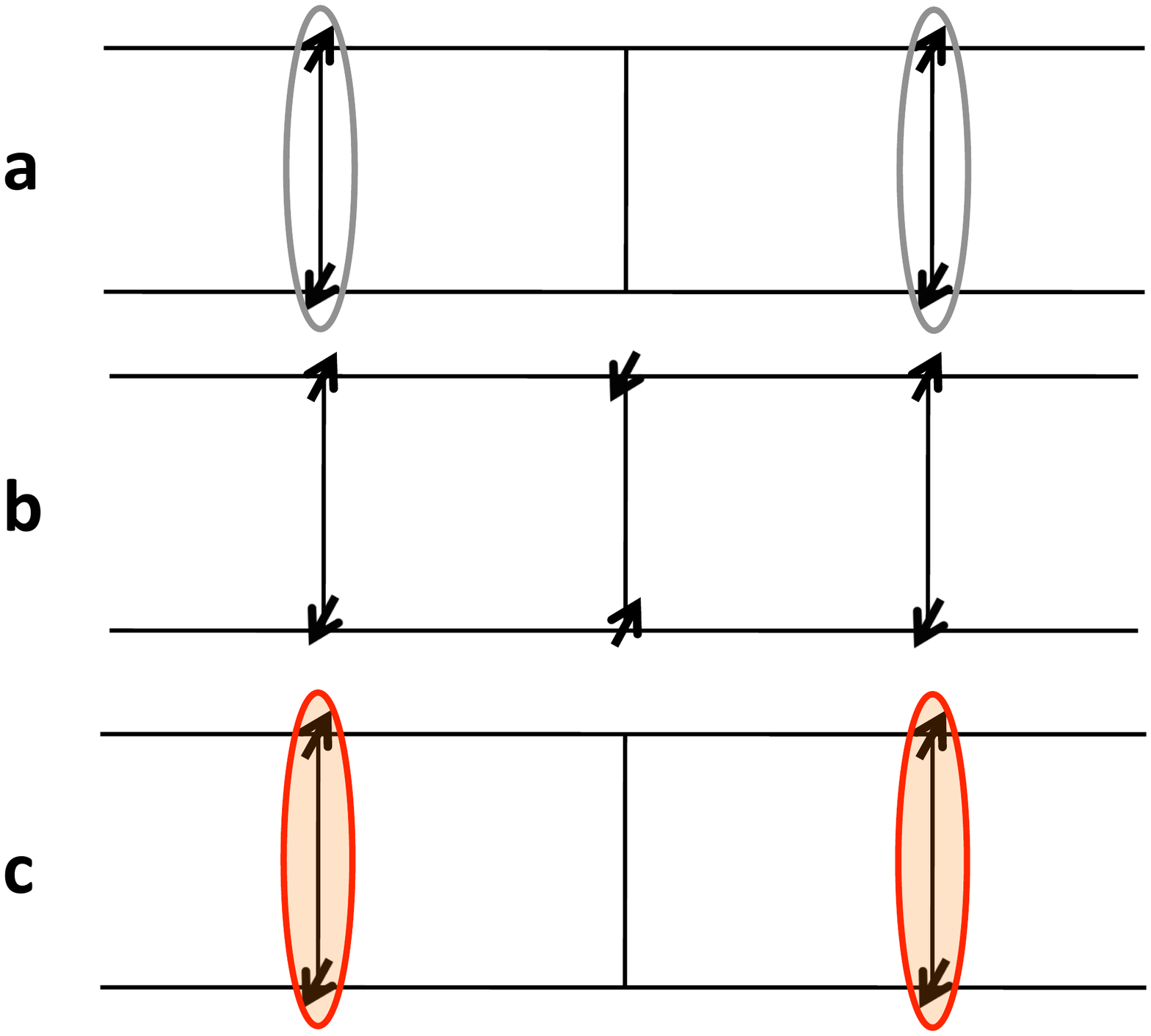}
  \caption{Possible phases for the repulsive Hubbard interaction, $U>0$:
 $d$-wave superfluid  (a), spin-density wave (b), spin-triplet
  superfluid along the $z$ direction (c).
The arrows and ellipses (shaded ellipses)
indicate the spins and spin-singlet pairing (spin-triplet pairing,
especially, $|S^z=0\rangle=
|\uparrow\downarrow\rangle+|\downarrow\uparrow\rangle$),
respectively. Since $U>0$, the on-site pairing is discouraged and
the interchain pairing is selected by the many-body effect for 
$t_{\perp\uparrow}\approx\pm t_{\perp\downarrow}$.
The SDW state realized 
has the alternate occupation in spin on the two legs.}
  \label{fig2}
 \end{center}
\end{figure}
\begin{figure}[h]
 \begin{center}
  \includegraphics[width=1.2\linewidth]{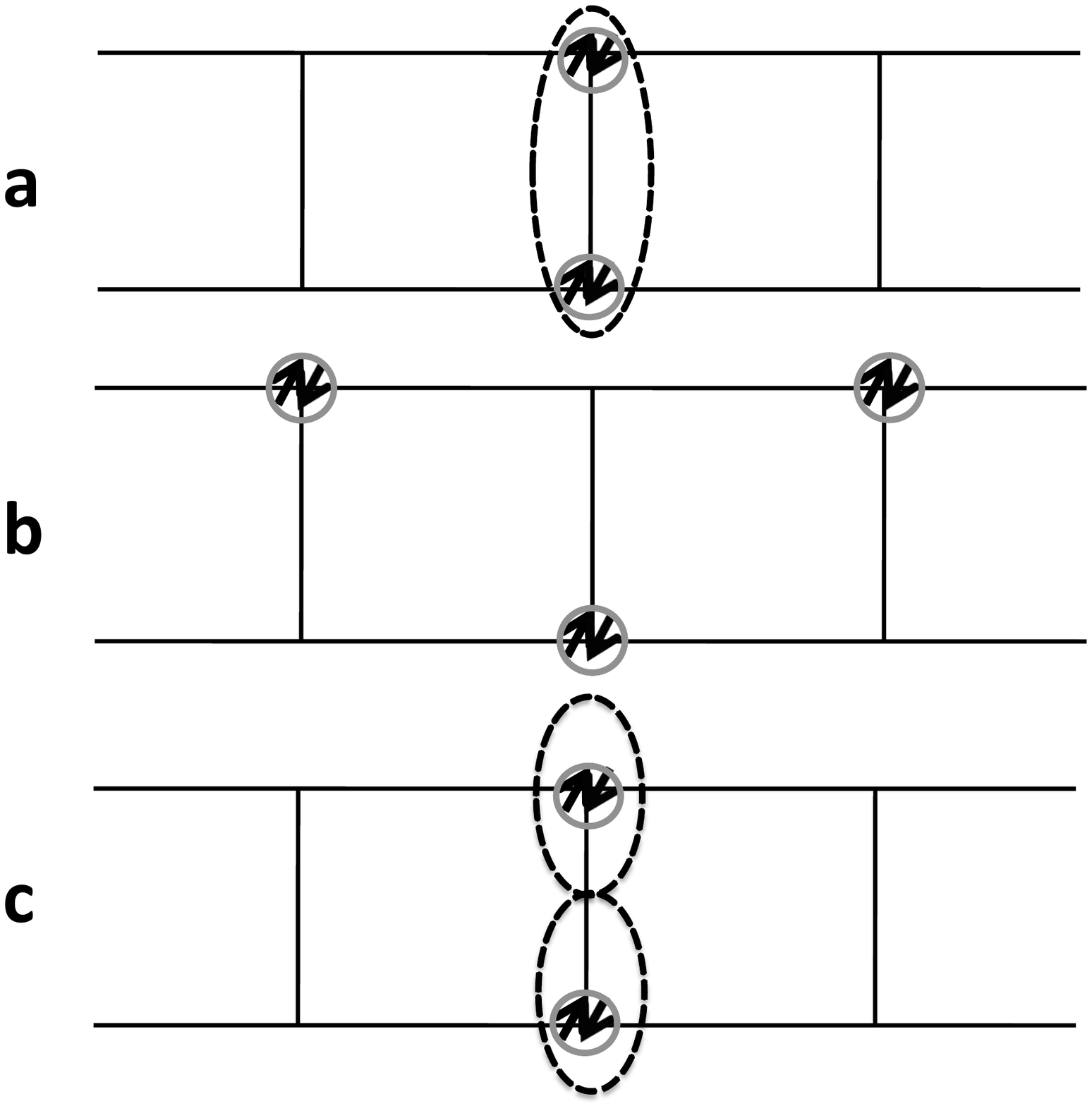}
  \caption{
  Possible phases for the attractive Hubbard interaction,
  $U<0$: bonding $s$-wave superfluid (a), charge density wave (b),
anti-bonding $s$-wave superfluid (c).
The difference of the dashed curves is that the
$s$-wave superfluid (a) occurs for the bonding band of the Cooper pairs
while the superfluid (c) occurs for the anti-bonding band of the 
Cooper pairs.
The CDW 
(b) has the alternate occupation on the two legs.}
  \label{fig3}
 \end{center}
\end{figure}

First, let us consider the case (a),
which corresponds to the limit $t_{\perp\uparrow}\approx t_{\perp\downarrow}$.
In this case, the terms proportional to
$g_2$, $g_5$ can be dropped due to the rapid oscillation of the cosines.
Thus, the RG equations reduce to ones without the state-dependent hopping
\cite{giamarchi2003quantum},
since this limit also allows us to do the substitutions,
$J_0(\delta_{\sigma-}\alpha)=1$ and $J_0(\delta_{\rho-}\alpha)=0$.
The RG analysis shows the fixed point is given by
$g_1\to-\infty$,
$g_3\to\infty$, $g_4\to\infty$, $g_6\to0$ for $U>0$ and
$g_1\to-\infty$, $g_3\to-\infty$, $g_4\to0$, $g_6\to\infty$ for $U<0$.
While regardless of the sign of the interaction,
$\phi_{\rho-}$, $\phi_{\sigma+}$, and $\phi_{\sigma-}$
are gapped, these minimums are different for opposite signs of the interaction.
It turns out that the minimum can be determined by the fixed point.
Then,
the dominant correlations are the $d$-wave superfluid for $U>0$
whose pairing occurs between the different chains
and the $s$-wave superfluid for $U<0$ whose
pairing essentially occurs in on-site.
The corresponding operators are
\beq
&&O_{\text{DSF}}(j)=\sum_{p}(
c_{j,\uparrow, p}c_{j,\downarrow, -p}-c_{j,\downarrow, p}c_{j,\uparrow, -p})
\nonumber\\
&&\sim e^{-i\theta_{\rho+}}(\cos\phi_{\rho-}\sin\phi_{\sigma+}
\sin\phi_{\sigma-}-i\sin\phi_{\rho}\cos_{\sigma+}\cos\phi_{\sigma-}),
\nonumber \\ \\
&&O_{\text{SSF}^{0}}(j)=\sum_{p}(
c_{j,\uparrow, p}c_{j,\downarrow, p}-c_{j,\downarrow, p}c_{j,\uparrow, p})
\nonumber\\
&&\sim e^{-i\theta_{\rho+}}(\cos\phi_{\rho-}
\cos\phi_{\sigma+}\cos\phi_{\sigma-}+i
\sin\phi_{\rho-}
\sin\phi_{\sigma+}\sin\phi_{\sigma-}),
\nonumber\\
\label{eq:s-wave-0}
\eeq
respectively \cite{giamarchi2003quantum}.
In contrast to the single chain Hubbard model,
we have for the ladder a superfluid regardless of sign of the interaction.

Let us next consider the case (b),
where both of the rung hoppings $t_{\perp\rho}\equiv t_{\perp\uparrow}
+t_{\perp\downarrow}$ 
and $t_{\perp\sigma}\equiv t_{\perp\uparrow}-t_{\perp\downarrow}$
are relevant and the substitutions
$J_0(\delta_{\rho-}\alpha)=J_0(\delta_{\sigma-}\alpha)=0$
are allowed.
In this case, the effects of $g_1$, $g_2$, $g_3$, $g_5$
can be dropped due to the large oscillations.
By solving the RG equations under these conditions,
the fixed points are shown to be
$g_4\to\infty$, $g_6\to\infty$ for $U>0$ and
$g_4\to-\infty$, $g_6\to-\infty$ for $U<0$.
Thus, we see $\theta_{\rho-}$, $\phi_{\sigma+}$, $\theta_{\sigma-}$
are going to be gapped.
From the fixed point analysis,
we find that
the following SDW and CDW operators
are relevant for $U>0$ and $U<0$, respectively:
\beq
&&O_{\text{SDW}^{\pi}}(j)=\sum_{p}
p( c^{\dagger}_{j,\uparrow,p}c_{j,\uparrow,p}
- c^{\dagger}_{j,\downarrow,p}c_{j,\downarrow,p}
)\nonumber\\
&&\sim e^{-i\phi_{\rho+}}(\sin\theta_{\rho-}
\cos\phi_{\sigma+}\cos\theta_{\sigma-}-
\cos\theta_{\rho-}\sin\phi_{\sigma+}\sin\theta_{\sigma-}
),\nonumber \\ \\
&&O_{\text{CDW}^{\pi}}(j)=
\sum_{p}p(c^{\dagger}_{j,\uparrow,p}c_{j,\uparrow,p}+
c^{\dagger}_{j,\downarrow,p}c_{j,\downarrow,p}
)\nonumber\\
&&\sim e^{-i\phi_{\rho+}}(\cos\theta_{\rho-}
\cos\phi_{\sigma+}\sin\theta_{\sigma-}
-\sin\theta_{\rho-}
\sin\phi_{\sigma+}\cos\theta_{\sigma-}),
\nonumber\\
\eeq
where $\pi$ indicates
the difference of the densities on the two legs.
The presence of the state-dependent rung hopping as
$|t_{\perp\uparrow}/t_{\perp\downarrow}|\ne1$ tries to destroy the 
the  superfluidity for the two-leg Hubbard ladder,
and causes fluctuations toward crystalline orders such as the SDW or
CDW.
Here, one may notice the analogy with
the single chain Hubbard system in the presence of a state-dependent hopping
\cite{PhysRevLett.95.226402} where
in the wide range of the parameters
SDW and CDW are shown to be the 
dominant fluctuations for $U>0$ and $U<0$, respectively.
Such emergences of the density wave states  in the single
chain system are natural, since one of the spin components
is reluctant to hop between different sites,
However, now we impose the spin dependence only for the rung direction.
Thus, the emergence of the SDW or CDW in our model is less trivial.

Let us finally consider the case (c), which can be realized
when $t_{\perp\uparrow}\approx -t_{\perp\downarrow}$ and
therefore the substitutions $J_0(\delta_{\rho-}\alpha)=1$
and $J_0(\delta_{\sigma-}\alpha)=0$ are justified.
In this case, $g_1$, $g_3$ can be dropped in a manner similar to
the other cases.
By solving the RG equations, we find the fixed points to
be
$g_2\to-\infty$, $g_4\to0$, $g_5\to\infty$, $g_6\to\infty$ for $U>0$
and
 $g_2\to-\infty$, $g_4\to0$, $g_5\to-\infty$, $g_6\to-\infty$ for $U<0$,
and therefore, $\phi_{\rho-}$, $\phi_{\sigma+}$, $\theta_{\sigma-}$
are gapped.
In accordance with the fixed points,
the dominant correlations are shown to be the spin-triplet superfluid
along the $z$ direction for $U>0$ and $s$-wave superfluid
for $U<0$, where
the corresponding operators are
\beq
&&O_{\text{TSF}^z}(j)=\sum_{p}
(c_{j,\uparrow,p}c_{j,\downarrow,-p}+
c_{j,\downarrow,p}c_{j,\uparrow,-p}
)\nonumber\\
&&\sim e^{-i\theta_{\rho+}}(\cos\phi_{\rho-}
\cos\phi_{\sigma+}\cos\theta_{\sigma-}-i\sin\phi_{\rho-}
\sin\phi_{\sigma+}\sin\theta_{\sigma-}),\nonumber \\ \\
&&O_{\text{SSF}^{\pi}}(j)=
\sum_{p}p(
c_{j,\uparrow,p}c_{j,\downarrow,p}-
c_{j,\downarrow,p}c_{j,\uparrow,p}
)\nonumber\\
&&\sim e^{-i\theta_{\rho+}}(
\sin\phi_{\rho-}\sin\phi_{\sigma+}\cos\theta_{\sigma-}+i
\cos\phi_{\rho-}\cos\phi_{\sigma+}\sin\theta_{\sigma-}),
\nonumber\\
\label{eq:s-wave-pi}
\eeq
respectively.
We first focus on the emergence of
the dominant fluctuation of the spin-triplet superfluid for $U>0$.
Namely, the sign inversion in the rung hopping
regarding only one of the spin components allows
the change of nature of the pairings from the inter-chain spin-singlet to
the inter-chain spin-triplet.
In the bonding and anti-bonding representation,
while the $d$-wave superfluid operator
has the form as $c_{j,\uparrow, 0}c_{j,\downarrow,0 }-c_{j,\uparrow,\pi
}c_{j,\downarrow,\pi}$,
the spin-triplet superfluid occurring is given as
 $c_{j,\uparrow, 0}c_{j,\downarrow,\pi }+c_{j,\downarrow,0
}c_{j,\uparrow,\pi}$.
To understand the mechanism, we first point out that
such a sign inversion in the rung hopping can be achieved by
introducing the Peierls phases both in charge and spin
sectors by $\pi/2$.
Then, what is important for the pairing is
the Peierls phase in the spin sector.
In fact, it has been shown
in  Ref.~\cite{PhysRevA.89.023623}
that such a Peierls phase causes the spin rotation
of the fermions for one of the chains and
transforms a spin-singlet into a spin-triplet pairing.
For $U<0$, on the other hand,
the difference between the $s$-wave superfluids in
Eq.~\eqref{eq:s-wave-pi} and in
Eq.~\eqref{eq:s-wave-0} is that if we treat the Cooper pairs
occurring in each chain as the bosons, 
the superfluid in the absence of the state-dependent hopping 
occurs for the bonding band
of the bosons while the superfluid in the presence of it occurs
for the anti-bonding band of the bosons.
Compared with the situation from the spin-singlet to
spin-triplet pairings for $U>0$,
the important ingredient for this change of the $s$-wave superfluids for $U<0$
is the Peierls phase in charge sector.
One may also accept this situation recalling that
in a Bose-Einstein condensate on a double well potential,
a BEC on the bonding band is normally the
ground state while a BEC on the anti-bonding band
becomes the ground state in the presence of the sign inversion hopping
\cite{pethick2002bose}.

The possible phases are summarized in Fig.~\ref{fig2} and Fig.~\ref{fig3}.

\section{Discussion}
\label{sec:discussion}
\subsection{Strong coupling limit}
So far, we have discussed the weak-coupling limit
by means of the bosonization and RG analysis,
it is also interesting to see what happens in the strong-coupling limit
in which naively a similar phase diagram may be expected.

For the $U>0$ case, in fact, it may be difficult to
depict a general phase diagram analytically
since a faithful effective Hamiltonian has yet to be known
except for commensurate filling such as half filling. In addition,
the rung hopping is a relevant perturbation, which
prevents one from starting at the single chain Hubbard model
where the Bethe ansatz approach is available.
At the same time, the previous numerical analyses
in the absence of the state-dependent hopping
show that the $d$-wave superfluid state emerges even in the
strong-coupling limit 
\cite{PhysRevB.53.11721,noack1996ground,PhysRevB.65.165122,PhysRevB.75.245119,
giamarchi2003quantum}.
In addition, since the hybridization among the four different Fermi
points by the on-site repulsive interaction
shown in Fig. \ref{fig1} (a) is an essential ingredient of the $d$-wave
state, we obtain the $d$-wave superfluid not only for 
$t_{\parallel}\gg t_{\perp}$ but also for $t_{\parallel}\approx
t_{\perp}$
in which the numerical calculation has been performed 
\cite{PhysRevB.53.11721,noack1996ground,PhysRevB.65.165122,PhysRevB.75.245119}.
Therefore, the presence of
the spin-triplet superfluid in the strong-coupling limit can
also be shown with the argument in Sec. \ref{sec:rg}.
Namely, 
by using the canonical transformations $c_{j,s,1}\to a_{j,s,1}$
$c_{j,s,-1}\to \pm a_{j,s,-1}$ where the sign is $+$ for $s=\uparrow$
and $-$ for $s=\downarrow$,
the Hamiltonian with $t_{\perp\uparrow}=-t_{\perp\downarrow}$
is mapped onto one with $t_{\perp\uparrow}=t_{\perp\downarrow}$,
that is, a normal two-leg fermionic Hubbard ladder can be obtained.
Accordingly, the operator of the spin-triplet superfluid is
transformed into that of the $d$-wave superfluid.
Therefore, once we confirm the emergence of the $d$-wave superfluid
in the normal two-leg fermionic Hubbard ladder system,
we see that the spin-triplet
superfluid occurring in $t_{\perp\uparrow}\approx -t_{\perp\downarrow}$
is robust.
We also note that the essence of the spin-triplet superfluid
is the manipulation on the rung hopping, and thus,
nothing happens and the $d$-wave superfluid remains
even if such a manipulation on the hopping is performed
for the chain direction.
Thus, to see the spin-triplet superfluid,
the manipulation on the hopping along the rung direction is required.
Another interesting but remaining issue may be the possibility
of segregation in the limit $t_{\uparrow}=0$ or
$t_{\downarrow}=0$
\cite{0305-4470-25-4-012,PhysRevLett.88.106401}.

On the other hand, for the $U<0$ case, we can discuss the possible phases in
the strong-coupling limit by means of
an effective Hamiltonian approach.
To see this, we first perform the so-called particle-hole transformation
\cite{Essler.etal/book.2010}
in this model. Then, the original model is mapped onto
the system with $U>0$ and spin imbalance at half filling, and therefore
the effective Hamiltonian is shown to be
\beq
H&=&J_{\parallel}\sum_{j}(\vec{S}_{j,1}\cdot
\vec{S}_{j+1,1}+\vec{S}_{j,-1}\cdot\vec{S}_{j+1,-1} )
-h\sum_{j}(S^z_{j,1}+S^z_{j,-1})\nonumber\\
&&+J^{xy}_{\perp}\sum_{j}(S^x_{j,1}S^x_{j,-1}+S^y_{j,1}S^y_{j,-1}).
+J^z_{\perp}\sum_{j}S^z_{j,1}S^z_{j,-1},
\eeq
where $J_{\parallel}=4t^2_{\parallel}/|U|$,
$J_{\perp}^{xy}=4t_{\perp\uparrow}t_{\perp\downarrow}/|U|$,
$J_{\perp}^z=2(t_{\perp\uparrow}^2+t_{\perp\downarrow}^2)/|U|$,
and $h$ is a magnetic field corresponding to filling in the original
attractive model.
By performing bosonization for the above Hamiltonian
\cite{giamarchi2003quantum},
one may obtain
\beq
H^{\text{eff}}=\sum_{\mu=s,a}\int \frac{dx}{2\pi}
\left(u_{\mu}K_{\mu}(\nabla\theta_{\mu})^2+\frac{u_{\mu}}{K_{\mu}}
(\nabla\phi_{\mu})^2
\right)\nonumber\\
+\frac{1}{2(\pi\alpha)^2}\int dx[J_{\perp}^{xy}\cos(\sqrt{2}\theta_a)
+ J_{\perp}^z\cos(2\sqrt{2}\phi_a)],
\label{eq:spin-h}
\eeq
where $\phi_{s(a)}=[\phi_1+(-)\phi_{-1}]/\sqrt{2}$
is the phase field in chain $p$, $\phi_p$ $(p=\pm1)$,
and similar relations for the $\theta$ field.
The original spin fields and phase fields are related as
$S^z_p(x)=-\nabla\phi_p(x)/\pi+(-1)^x\cos(2\phi_p(x))/(\pi\alpha)$
and  $S^+_p(x)=e^{-i\theta_p(x)}[(-1)^x+\cos2\phi_p(x)]/\sqrt{2\pi\alpha}$.
Since $J_{\parallel} \gg J_{\perp}$ is concerned, we can determine the
Tomonaga-Luttinger
parameters as
\beq
K_{s,a}=K\left(1\mp \frac{KJ_{\perp}^z}{2\pi u}\right).
\label{eq:k-spin}
\eeq
Here $K$ and $u$ are the Tomonaga-Luttinger parameter and velocity in
the single chain Heisenberg model, respectively.
The Tomonaga-Luttinger parameter $K$ can be determined
by means of Bethe ansatz, and it is known that the possible range is
$1/2\le K\le 1$, where
$K=1/2$ corresponds to the no magnetization case and
$K=1$ to the fully polarized case \cite{PhysRevLett.45.1358}.
Since the above consists of the linear combination of the simple cosine
terms, one can determine the ground state with a simple scaling argument.
In fact, $\cos\sqrt{2}\theta_a$ and $\cos\sqrt{8}\phi_a$
have the scaling dimensions of $(2K_a)^{-1}$ and $2K_a$,
respectively.
Thus, we see that $\cos\sqrt{2}\theta_a$ is ordered for $K_a>1/2$
and the situation is reversed for $K_{a}<1/2$.
As can be seen from Eqs. \eqref{eq:spin-h} and \eqref{eq:k-spin},
$K_a>1/2$, and we expect that $\theta_a$ is ordered except for the limit
$J_{\perp}^{xy}\to0$ where $\phi_a$ is ordered.
To specify the ground state in the spin language, let us
introduce bonding and anti-bonding spin operators
as $\vec{S}_0=\vec{S}_1+\vec{S}_{-1}$ and
$\vec{S}_{\pi}=\vec{S}_1-\vec{S}_{-1}$,
respectively.
Then, one finds that the bonding (anti-bonding)
transverse spin-spin correlation
$\langle S^+_{0}(r)S^-_{0}(0)\rangle$
$(\langle S^+_{\pi}(r)S^-_{\pi}(0)\rangle)$
 is dominant
for $\theta_a$ to be gapped with $J_{\perp}^{xy}<0$
$(J_{\perp}^{xy}>0)$,
while
the anti-bonding longitudinal spin-spin correlation
$\langle S_{\pi}^z(r)S_{\pi}^z(0)\rangle$ is dominant for $\phi_a$
to be gapped \cite{giamarchi2003quantum}.
Now, we can determine the dominant correlation in the original model
by using the particle-hole transformation again.
Since by this transformation
\beq
S^{-}_{0}\to\sum_{p}pc_{j,\uparrow,p}c_{j,\downarrow,p}
=O_{\text{SSC}^{\pi}},\\
S^{-}_{\pi}\to\sum_{p} c_{j,\uparrow,p}c_{j,\downarrow,p}
= O_{\text{SSC}^0},\\
S^z_{\pi}\to\sum_{p,s}p c^{\dagger}_{j,s,p}c_{j,s,p}
=O_{\text{CDW}^{\pi}},
\eeq
we conclude that the $s$-wave superfluid is dominant except for
$t_{\perp\uparrow}t_{\perp\downarrow}\to0$ where
the CDW correlation is dominant.
In particular, the bonding $s$-wave pairing state
 is realized
for $t_{\perp\uparrow}t_{\perp\downarrow}>0$ while
the anti-bonding $s$-wave pairing state is realized for the opposite
sign case.
Thus, the phase structure is compatible with
the weak-coupling analysis while
in the weak-coupling limit
the region of the $s$-wave superfluid is rather narrow but
in the strong-coupling limit the situation is reversed.
This may be explained by the observation that
the pairing gap becomes larger as the attractive interaction
is increased
and the pairing in the $s$-wave superfluid essentially occurs
in a single site, and therefore the introduction of the small state-dependent
rung hopping may not cause the disappearance of the superfluid correlation.

\subsection{Experimental protocol}

We now discuss the realization of our model and its
ground states in cold atoms.

In order to realize the two-leg ladder geometry, 
we can consider an optical superlattice
\cite{PhysRevA.73.033605,folling2007direct}.
By using this technology, we can obtain a system
where there are a number of two-leg ladders, each of which
is weakly coupled by some hopping parameter.
To ensure the one-dimensional character in the system,
this hopping parameter should be much smaller than
a temperature \cite{giamarchi2003quantum}.
Then, the two-leg ladder system in the absence of
state-dependent hopping is obtained. 
As another route to realize such a ladder geometry,
one may also utilize internal degrees of freedom in an atom 
as discussed in Refs. \cite{gorshkov2010two,PhysRevLett.112.043001}.

On the other hand, 
the Hubbard interaction $U$ can be tuned by selecting atomic species
and by changing lattice depth.
Typically, $^{40}$K and $^{6}$Li
have been utilized to realize the system for $U>0$ and $U<0$,
respectively \cite{RevModPhys.80.885}.
In addition, the Feshbach resonance is available to change
the strength and sign of the interaction.

The most important ingredient in the system discussed
is a state-dependent hopping.
When it comes to a positive state-dependent hopping,
heteronuclear mixtures such as $^{6}$Li$-^{40}$K
\cite{PhysRevLett.100.053201} and  $^{6}$Li$-^{173}$Yb
\cite{PhysRevLett.106.205304}
are available.
However, since we are also interested in
a state-dependent  hopping whose sign is different between spin up and
down,
another scheme is necessary.

To this end, we start with the two-leg Hubbard ladder system
in the absence of a state-dependent hopping.
To obtain a state-dependent rung hopping, 
we consider adding the following time-dependent term
in the Hamiltonian:
\beq
\sum_{s=\uparrow,\downarrow}A_{s}\cos(\omega t)
\sum_{p=\pm1}pn_{j,s,p}.
\eeq
If $A_{\uparrow}=A_{\downarrow}$,
the above time-varying linear potential can be obtained with a
sinusoidal shaking of an optical lattice along the rung direction
\cite{PhysRevLett.99.220403}. 
In order to exactly obtain the above time-dependent term for 
$A_{\uparrow}\ne A_{\downarrow}$, a sinusoidal shaking of
a state-dependent optical lattice 
\cite{PhysRevLett.91.010407,mandel2003controlled}
or of a magnetic
field gradient \cite{PhysRevLett.111.185301}
 can be utilized.
Then, an essential point is that when  $\hbar\omega\gg 
t_{\parallel},t_{\perp}, |U|$,
we may perform the time average
of the above oscillation term,
which causes the renormalization of the hopping parameter as
$t_{\perp}\to t_{\perp}J_0(A_{s}/(\hbar\omega))$
\cite{PhysRevLett.95.260404,PhysRevX.4.031027}.
Since the argument of the Bessel function is now state dependent,
the Hamiltonian \eqref{eq:original-h} can be obtained.
We note that the Bessel function can take a negative
value, which allows us to consider a negative hopping parameter.
Indeed, such a negative hopping parameter by the time-dependent oscillation term
has been observed in Refs.~\cite{PhysRevLett.99.220403,struck2011quantum}

A particularly interesting challenge is to
make the spin-triplet superfluid realized
around $t_{\perp\uparrow}\approx -t_{\perp\downarrow}$
for a repulsive $U$.
Here, we note that  $''\approx''$ implies
$t_{\perp\rho}/T\ll 1$ where $T$ is a temperature.
In this case, the effect of $t_{\perp\rho}$ can be dropped in the Hamiltonian,
and the effective Hamiltonian reduces to one of
$t_{\perp\uparrow}=-t_{\perp\downarrow}$
\cite{giamarchi2003quantum}.
In the two-leg fermionic ladder system at incommensurate filling,
we have a one charge gap and two spin gaps, 
which are exponentially small
for the weak-coupling limit but are of the order of the exchange energy
for the strong-coupling limit (
See Refs.
\cite{PhysRevB.53.11721,noack1996ground,PhysRevB.65.165122,PhysRevB.75.245119}
for numerical estimations for the strong-coupling limit).
Since the gaps are essential to characterize the spin-triplet superfluid state,
the temperature should be smaller than them as well as the
other Hamiltonian parameters $t_{\parallel},U,t_{\perp,s}$.
Thus, the dominant spin-triplet superfluid correlation
should show up at the temperature satisfying these conditions.
We note that
a similar argument for the realizations of the other phases
is also possible.

When it comes to the spin-triplet superfluid,
we can also utilize the technique of synthetic gauge fields
\cite{RevModPhys.83.1523}.
Recently, by using a Raman laser and lattice driving
\cite{PhysRevLett.107.255301,PhysRevLett.108.225303,PhysRevLett.108.225304,
PhysRevLett.111.185301,PhysRevLett.111.185302},
it is possible to introduce the Peierls phase in the hopping parameter.
If such a Peierls phase has a state dependency, which is indeed
possible experimentally,
the hopping parameter is modified as $t\to te^{i\Phi_{s}}$,
where $\Phi_{s}$ is the Peierls phase.
When $\Phi_{\uparrow}=-\Phi_{\downarrow}$, such a hopping term
can also be regarded as a spin-orbit coupling.
As explained in Sec. \ref{sec:rg}, the essence of the spin-triplet superfluid
is the emergence of the state-dependent Peierls phase 
along the rung direction as
$t_{\perp}\to t_{\perp }e^{is\Phi}$.
Thus, the spin-triplet superfluid realized with this manipulation
is essentially the same mechanism as one discussed in this paper.

Finally, we give a few comments on the experimental observability
of the superfluids. 
An important feature is the presence of
the gaps, which may be measured by rf spectroscopy 
\cite{stewart2008using}.
However, 
the measurement of  gaps alone is not enough to distinguish
different superfluids.
One of the possible solutions to this problem is 
to use the particle-hole transformation \cite{Essler.etal/book.2010}.
Then, the Hamiltonian for $U>0$ 
 at incommensurate filling without a spin imbalance 
is transformed into one for $U<0$ at half filling with a spin imbalance.
We also note that  the hopping terms in Eq. \eqref{eq:original-h}
are invariant under such a particle-hole 
transformation.
On the other hand, 
the operators of the $d$-wave and spin-triplet 
superfluids are transformed into those of
staggered spin-flux and bond SDW phases, respectively
\cite{PhysRevA.89.023623}.
The properties of such phases may be captured by
the local addressing of the flux 
\cite{PhysRevLett.107.255301,aidelsburger2013experimental}
and spin correlations \cite{PhysRevLett.105.265303,greif2013short}
or by spin-sensitive Bragg scattering of light 
\cite{hart2014observation}.

\section{conclusion}
\label{sec:conclusion}
We have examined a two-leg fermionic Hubbard ladder model
in the presence of a state-dependent hopping.
We have focused on a case where such a hopping exists for the rung
direction. This system can be treated as the minimal construction of
the physics of mixed dimensions \cite{PhysRevLett.101.170401}
if the rung hopping in one of the states (spins)
is zero 
since another rung hopping plays a role in connecting
different chains.
Due to the state-dependent hopping,
the original $d$-wave and $s$-wave superfluids realized in
the normal two-leg fermionic Hubbard ladder model
for the repulsive
and attractive interactions, respectively
become unstable.
We have demonstrated that
instead the spin-triplet superfluid, SDW, and CDW states become
stable depending on the ratio $t_{\perp\uparrow}/t_{\perp\downarrow}$.
In particular, our proposal shows a spin-triplet superfluid 
state for purely local interactions can be realized. 
We have also discussed the experimental protocol and
observability toward the spin-triplet superfluid.

\section*{acknowledgement}
SU thanks A. Tokuno for fruitful discussions on the realization
of our model.
This work was supported by the Swiss National Foundation under division II.

\appendix

\section*{Appendix: Renormalization group equations}
In this Appendix, we wish to outline
the derivation of the renormalization group equations
in a similar way as Ref.~\cite{giamarchi2003quantum}.
We first consider the following correlation function:
\beq
R(r_1-r_2)=\langle T_{\tau}e^{i\phi_{\sigma+}(x_1,\tau_1)}e^{-i\phi_{\sigma+}
(x_2,\tau_2)} \rangle
\eeq
where $T_{\tau}$ denotes the time-ordered product.
By expanding the above correlation function in terms of $g_i$
up to third order, we obtain
\beq
R(r_1-r_2)\approx
e^{-\frac{K_{\sigma+}}{2}F_1(r_1-r_2)}
+(\text{S})+(\text{T}),
\eeq
where $F_1(r)=\ln(r/\alpha)$,
\begin{widetext}
\beq
(\text{S})=\frac{1}{2}\left(\frac{g_1}{8(\pi\alpha)^2v_F}\right)^2
\sum_{\epsilon_1,\epsilon_2=\pm1}\int d^2r'd^2r''
\Big[\langle
e^{i[\phi_{\sigma+}(\vec{r}_1)-\phi_{\sigma+}(\vec{r}_2)+
2\epsilon_1(\phi_{\sigma+}(\vec{r}')
-\phi_{\sigma+}(\vec{r}''))
+2\epsilon_2(\phi_{\sigma-}(\vec{r}')-\phi_{\sigma-}(\vec{r}'')
-\delta_{\sigma-}(x'-x''))
]}
\rangle_0
\nonumber\\
-\langle
e^{i(\phi_{\sigma+}(\vec{r}_1)-\phi_{\sigma+}(\vec{r}_2))}\rangle_0
\langle^{i[
2\epsilon_1(\phi_{\sigma+}(\vec{r}')
-\phi_{\sigma+}(\vec{r}''))
+2\epsilon_2(\phi_{\sigma-}(\vec{r}')-\phi_{\sigma-}(\vec{r}'')
-\delta_{\sigma-}(x'-x''))
]}
\rangle_0
\Big]\nonumber\\
+\frac{1}{2}\left(\frac{g_2}{8(\pi\alpha)^2v_F}\right)^2
\sum_{\epsilon_1,\epsilon_2=\pm1}\int d^2r'd^2r''
\Big[\langle
e^{i[\phi_{\sigma+}(\vec{r}_1)-\phi_{\sigma+}(\vec{r}_2)+
2\epsilon_1(\phi_{\sigma+}(\vec{r}')
-\phi_{\sigma+}(\vec{r}''))
+2\epsilon_2(\phi_{\rho-}(\vec{r}')-\phi_{\rho-}(\vec{r}'')
-\delta_{\rho-}(x'-x''))
]}
\rangle_0
\nonumber\\
-\langle
e^{i(\phi_{\sigma+}(\vec{r}_1)-\phi_{\sigma+}(\vec{r}_2))}\rangle_0
\langle^{i[
2\epsilon_1(\phi_{\sigma+}(\vec{r}')
-\phi_{\sigma+}(\vec{r}''))
+2\epsilon_2(\phi_{\rho-}(\vec{r}')-\phi_{\rho-}(\vec{r}'')
-\delta_{\rho-}(x'-x''))
]}
\rangle_0
\Big]\nonumber\\
+\frac{1}{2}\left(\frac{g_4}{8(\pi\alpha)^2v_F}\right)^2
\sum_{\epsilon_1,\epsilon_2=\pm1}\int d^2r'd^2r''
\Big[\langle
e^{i[\phi_{\sigma+}(\vec{r}_1)-\phi_{\sigma+}(\vec{r}_2)+
2\epsilon_1(\phi_{\sigma+}(\vec{r}')
-\phi_{\sigma+}(\vec{r}''))
+2\epsilon_2(\theta_{\rho-}(\vec{r}')-\theta_{\rho-}(\vec{r}''))
]}
\rangle_0
\nonumber\\
-\langle
e^{i(\phi_{\sigma+}(\vec{r}_1)-\phi_{\sigma+}(\vec{r}_2))}\rangle_0
\langle^{i[
2\epsilon_1(\phi_{\sigma+}(\vec{r}')
-\phi_{\sigma+}(\vec{r}''))
+2\epsilon_2(\theta_{\rho-}(\vec{r}')-\theta_{\rho-}(\vec{r}''))
]}
\rangle_0
\Big]
\nonumber\\
+
\frac{1}{2}\left(\frac{g_6}{8(\pi\alpha)^2v_F}\right)^2
\sum_{\epsilon_1,\epsilon_2=\pm1}\int d^2r'd^2r''
\Big[\langle
e^{i[\phi_{\sigma+}(\vec{r}_1)-\phi_{\sigma+}(\vec{r}_2)+
2\epsilon_1(\phi_{\sigma+}(\vec{r}')
-\phi_{\sigma+}(\vec{r}''))
+2\epsilon_2(\theta_{\sigma-}(\vec{r}')-\theta_{\sigma-}(\vec{r}''))
]}
\rangle_0
\nonumber\\
-\langle
e^{i(\phi_{\sigma+}(\vec{r}_1)-\phi_{\sigma+}(\vec{r}_2))}\rangle_0
\langle^{i[
2\epsilon_1(\phi_{\sigma+}(\vec{r}')
-\phi_{\sigma+}(\vec{r}''))
+2\epsilon_2(\theta_{\sigma-}(\vec{r}')-\theta_{\sigma-}(\vec{r}''))
]}
\rangle_0
\Big],
\eeq
and
\beq
(\text{T})&=&-g_1g_3g_4\left(\frac{1}{8(\pi\alpha)^2v_F}\right)^3
\sum_{\epsilon_1,\epsilon_2,\epsilon_3=\pm1}
\int d^2r'd^2r''d^2r'''\nonumber\\
&&\Big[
\langle
e^{i[\phi_{\sigma+}(\vec{r}_1)-\phi_{\sigma+}(\vec{r}_2)+
2\epsilon_1(\phi_{\sigma+}(\vec{r}')
-\phi_{\sigma+}(\vec{r}'''))
+2\epsilon_2(\phi_{\sigma-}(\vec{r}')-\phi_{\sigma-}(\vec{r}'')
-\delta_{\sigma-}(x'-x''))
+2\epsilon_3(\theta_{\rho-}(\vec{r}'')-\theta_{\rho-}(\vec{r}'''))
]}
\rangle_0
\nonumber\\
&&-\langle
e^{i(\phi_{\sigma+}(\vec{r}_1)-\phi_{\sigma+}(\vec{r}_2))}\rangle_0
\langle^{i[
2\epsilon_1(\phi_{\sigma+}(\vec{r}')
-\phi_{\sigma+}(\vec{r}'''))
+2\epsilon_2(\phi_{\sigma-}(\vec{r}')-\phi_{\sigma-}(\vec{r}'')
-\delta_{\sigma-}(x'-x''))
+2\epsilon_3(\theta_{\rho-}(\vec{r}'')-\theta_{\rho-}(\vec{r}'''))]}
\rangle_0
\Big]\nonumber\\
&&-g_2g_5g_6\left(\frac{1}{8(\pi\alpha)^2v_F}\right)^3
\sum_{\epsilon_1,\epsilon_2,\epsilon_3=\pm1}
\int d^2r'd^2r''d^2r'''\nonumber\\
&&\Big[
\langle
e^{i[\phi_{\sigma+}(\vec{r}_1)-\phi_{\sigma+}(\vec{r}_2)+
2\epsilon_1(\phi_{\sigma+}(\vec{r}')
-\phi_{\sigma+}(\vec{r}''))
+2\epsilon_2(\phi_{\rho-}(\vec{r}')-\phi_{\rho-}(\vec{r}''')
-\delta_{\rho-}(x'-x'''))
+2\epsilon_3(\theta_{\rho-}(\vec{r}'')-\theta_{\rho-}(\vec{r}'''))
]}
\rangle_0
\nonumber\\
&&-\langle
e^{i(\phi_{\sigma+}(\vec{r}_1)-\phi_{\sigma+}(\vec{r}_2))}\rangle_0
\langle^{i[
2\epsilon_1(\phi_{\sigma+}(\vec{r}')
-\phi_{\sigma+}(\vec{r}''))
+2\epsilon_2(\phi_{\rho-}(\vec{r}')-\phi_{\rho-}(\vec{r}''')
-\delta_{\rho-}(x'-x'''))
+2\epsilon_3(\theta_{\rho-}(\vec{r}'')-\theta_{\rho-}(\vec{r}'''))]}
\rangle_0
\Big].
\eeq
In the above, $\langle\cdots\rangle_0$ denotes the average without
the cosine terms, that is, one with the Tomonaga-Luttinger
Hamiltonian.
When we focus on $(\text{T})$,
the dominant contributions come from
$\vec{r}'''=\vec{r}''+\vec{r}$ or $\vec{r}''=\vec{r}'+\vec{r}$
for the term proportional to $g_1g_3g_4$
and from
$\vec{r}'''=\vec{r}''+\vec{r}$ or $\vec{r}'''=\vec{r}'+\vec{r}$
for one proportional to $g_2g_5g_6$
with a small $r$.
Therefore, by expanding around $\vec{r}=0$,
after a straightforward calculation,
we can obtain the following renormalization relations on
the effective quantities:
\beq
&&K_{\sigma+}^{\text{eff}}=
K_{\sigma+}-\frac{K_{\sigma+}^2}{2}\int \frac{dr}{\alpha}
\Big[y_1^2
\left(\frac{r}{\alpha}\right)^{3-2(K_{\sigma+}+K_{\sigma-})}J_0
(2\delta_{\sigma-}r)
+y_2^2\left(\frac{r}{\alpha}\right)^{3-2(K_{\sigma+}+K_{\rho-})}J_0
(2\delta_{\rho-}r)\nonumber\\
&&+y_4^2\left(\frac{r}{\alpha}\right)^{3-2(K_{\sigma+}+1/K_{\rho-})}
+y_6^2\left(\frac{r}{\alpha}\right)^{3-2(K_{\sigma+}+1/K_{\sigma-})}
\Big],\\
&&(y_1^{\text{eff}})^2=y_1^2-2y_1y_3y_4\int \frac{r}{\alpha}
\left(\frac{r}{\alpha}\right)^{1-2/K_{\rho-}},\\
&&(y_2^{\text{eff}})^2=y_2^2-2y_2y_5y_6\int \frac{r}{\alpha}
\left(\frac{r}{\alpha}\right)^{1-2/K_{\sigma-}},\\
&&(y_4^{\text{eff}})^2=y_4^2-2y_1y_3y_4\int \frac{r}{\alpha}
\left(\frac{r}{\alpha}\right)^{1-2K_{\rho-}}J_0(2\delta_{\sigma-}r),\\
&&(y_6^{\text{eff}})^2=y_6^2-2y_2y_5y_6\int \frac{r}{\alpha}
\left(\frac{r}{\alpha}\right)^{1-2K_{\rho-}}J_0(2\delta_{\rho-}r).
\eeq
By changing the cutoff $\alpha\to e^l\alpha=\alpha+d\alpha$,
we obtain Eqs. (10), (13), (14), (16), and (18).
In a way similar to the above, the other RG equations can also be obtained.

\end{widetext}

%

\end{document}